\documentstyle[aps,prb]{revtex}
\begin{document}
\title{
Addition energies in semiconductor quantum dots: \\
Role of electron-electron interaction
}
\author{Massimo Rontani, Fausto Rossi, Franca Manghi, and Elisa Molinari}
\address{
Istituto Nazionale per la Fisica della Materia (INFM), and
Dipartimento di Fisica, \\
Universit\`a di Modena, via Campi 213/A, I-41100 Modena, Italy
}
\date{\today}
\maketitle
\begin{abstract}

We show that the addition spectra of semiconductor quantum dots in the
presence of magnetic field can be studied through a theoretical scheme that
allows an accurate and practical treatment of the single particle states
and electron-electron interaction up to large numbers of electrons.
The calculated addition spectra exhibit the typical structures of Hund-like
shell filling, and account for recent experimental findings.
A full three dimensional description of Coulomb interaction is found to be
essential for predicting the conductance characteristics of few-electron
semiconductor structures.

\end{abstract}

In semiconductor quantum dots (QDs) Coulomb correlation effects are
expected to become more and more pronounced as their spatial confinement is
increased. This has been sofar one of the reasons of interest for such 
systems: they display the rich shell structure typical of atomic physics,
but in addition they offer the possibility to tailor the confining
potential and to vary the electron occupation by adjusting external
parameters,\cite{review_dots} thus allowing direct investigation of
fundamental properties of charge correlation.

At the same time, however, these results imply that many body effects
cannot be neglected in the theoretical description of QDs, and may be very
relevant for predictions on their possible applications, for example
in semiconductor-based few-electron devices.
While capacitance and tunneling spectroscopy experiments are becoming
available for different dot structures,
\cite{Ashoori,McEuen,Schmidt,Tarucha,Fricke} it is still an open and
difficult problem to devise theoretical approaches allowing to include
many body effects in calculations for realistic systems.
Exact calculations\cite{exact} are necessarily limited to very few
electrons; state-of-the-art QD structures instead involve several electrons
(of the order of ten or hundred), therefore in general out of reach for
these methods.

In this letter we propose an accurate and effective theoretical scheme that
includes electron-electron interaction and can be applied to systems up to
a large number of electrons. This approach can treat arbitrary 
three dimensional (3D) confinement potentials, and hence deal
with realistic QD structures also in the presence of external magnetic
field. We apply this scheme to structures that were investigated in a
recent experiment. The results are in good quantitative agreement with the
observed spectra,\cite{Tarucha} and provide a transparent interpretation of
their physical origin.

A key quantity that characterizes transport into a QD is the addition
energy, i.e.~the energy $A\!\left(N\right)$ required in order to place an
extra electron into a dot that is initially occupied by $N-1$ particles.
Such quantity, analogous to electron affinity in atomic physics, can be
measured experimentally as a function of $N$.
A systematic experiment on disk-shaped dots\cite{Tarucha} has shown that
the voltage increment $\Delta A$ between successive single electron
tunneling processes - i.e.~between two successive maxima in the
conductance - peaks at``magic'' values of $N$
($N = 2, 6, 12$), consistent with the filling of complete shells
calculated\cite{Macucci} for a dot of the same symmetry. Moreover,
the experimental analysis showed that unusually large values of
$\Delta A$ also occur for values of $N$ corresponding to half-shell filling
($N = 4, 9$).
The origin of these features,
reminiscent of Hund's rule in atomic physics,\cite{Tarucha,Wang,Wojs} is
intimately related to electron-electron interaction; therefore their
quantitative evaluation is a challenge for any theoretical analysis of
addition spectra in QDs.

Our description is based on an accurate treatment of the single
particle Hamiltonian and on the inclusion of Coulomb correlation according
to first order perturbation theory: assuming the many particle ground state
to be described by a single Slater determinant, the total energy of the
full interacting Hamiltonian is 
\begin{equation}
E\!\left(N\right)
= \sum_{\alpha \sigma} \varepsilon_{\alpha} \langle \hat{n}_{\alpha
\sigma} \rangle + \frac{1}{2}
      \sum_{\alpha \beta \sigma} \left[U_{\alpha \beta}
\langle \hat{n}_{\beta -\sigma} \rangle +
      (U_{\alpha \beta}-J_{\alpha \beta}) \langle
       \hat{n}_{\beta \sigma} \rangle\right]
      \langle \hat{n}_{\alpha \sigma} \rangle ,
\label{eq:E}
\end{equation}
where $\langle \hat{n}_{\alpha \sigma} \rangle$ is the orbital occupation
number, $\alpha$ denoting the set of radial and angular
quantum numbers $(n,m)$ and $\sigma$ the spin value.
According to (\ref{eq:E}), the evaluation of the energies for the various
many electron states in the dot requires only the knowledge
of the Coulomb and exchange matrix elements
\begin{equation}
U_{\alpha\beta} = e^2\int\!\!\!\int
\frac{ {\left|{\phi}_{\alpha}\!\left(\bbox{r}\right)\right|}^2
{\left|{\phi}_{\beta}\!\left(\bbox{r^{\prime}}\right)\right|}^2}
{\kappa\left|\bbox{r}-\bbox{r^{\prime}}\right|}
{\rm\,d}\bbox{r}{\rm\,d}\bbox{r^{\prime}} ,
\label{eq:Coulomb}
\end{equation}
\begin{equation}
J_{\alpha\beta} = e^2\int\!\!\!\int
\frac{ {{\phi}_{\alpha}}^{\!*}\!\left( \bbox{r} \right)
{{\phi}_{\beta}}^{\!*}\!\left(\bbox{r^{\prime}}\right)
{\phi}_{\alpha}\left(\bbox{r^{\prime}}\right)
{\phi}_{\beta}\left(\bbox{r}\right) }{\kappa\left|\bbox{r}-
\bbox{r^{\prime}}
\right|}
{\rm\,d}\bbox{r}{\rm\,d}\bbox{r^{\prime}}\ .
\label{eq:exchange}
\end{equation}
Here $\kappa$ is the dielectric constant and $\phi_\alpha$ are
single particle real-space eigenstates of the dot in the presence of the
applied magnetic field.

The important feature of the present scheme is that $U_{\alpha\beta}$,
$J_{\alpha\beta}$ are calculated directly, contrary to other common 
approaches where Coulomb and exchange integrals are taken as input 
parameters.\cite{manghi} 
Thus, many body effects are taken into account through (\ref{eq:E}),
and quantum-confinement effects through the single particle
ingredients, energies $\varepsilon_\alpha$ and wavefunctions $\phi_\alpha$;
the latter enter only in the calculation of Coulomb and exchange matrix
elements.
This has the advantage that realistic confining potentials
(e.g.~extracted from high-resolution TEM micrographs) and an external
magnetic field can be included quite easily by chosing the appropriate
single particle Hamiltonian, as is now commonly done for the calculation
of optical properties.\cite{noi_excitons}

Note that previous approaches have modeled the QD confinement in terms of a
purely two dimensional (2D) coordinate space.\cite{exact,Wang,Wojs}
As a matter of fact, in most state-of-the-art QD structures the
confinement is far from being 2D: for example, in the gated dots of Ref.
\onlinecite{Tarucha} the typical confinement width in the growth direction
$z$ (of the order of 10nm) is comparable with the typical extension of the
carrier ground state in the lateral-confinement ($xy$) plane.
Therefore, in our calculations we fully retain the 3D nature of the problem
and mimic the dots of Ref. \onlinecite{Tarucha}
by assuming a finite-barrier quantum-well
potential along $z$ and a parabolic potential in ($xy$).
Note that the assumption of parabolic potentials reflects the experimental
indications due to the specific sample structure and is {\em not} required
by our model.
Within the usual envelope-function approximation, the only input parameters
are the electron effective masses ($m^{\ast} = 0.065m_0$ in the dot and
$m^{\ast}=0.079m_0$ in the barrier), the dielectric constant in the dot
($\kappa = 12.98$), and the quantum-well height and width ($V_0
= 200$~meV and $L = 12$~nm).

Our results for the addition energy variations,
$\Delta A\!\left(N\right) =
A\!\left(N+1\right)-A\!\left(N\right)$,
are displayed in Fig.~\ref{fig1} as a function of the
electron number $N$ for different dots,
characterized by different in-plane confinement
energies $\hbar\omega_0$. Here,
$A\!\left(N\right)$ is obtained as
$E\!\left(N\right)-E\!\left(N-1\right)$,
where $E\!\left(N\right)$ is the ground state energy in Eq.~(\ref{eq:E}).
As we can see,
$\Delta A\!\left(N\right)$ exhibits
peaks corresponding both to complete and half shell filling,
and is in excellent agreement with experiments in 
Ref.~\onlinecite{Tarucha}.

The origin of this result is in the delicate
interplay between single particle contributions and electron-electron
repulsion, which according to Eq.~(\ref{eq:E}) defines the ground state
configuration: the single particle term favours complete shell filling,
while the repulsion among parallel-spin electrons, smaller than the
repulsion among opposite-spin ones, makes the configurations with maximum
total spin energetically favoured. Hund's rule is therefore already
contained in Eq.~(\ref{eq:E}) and clearly explains the physical origin of
the half shell-filling
structures. In fact, adding an electron to a half-filled shell forces
the double occupancy of a level: consequently,
$\Delta A$ is raised by the dominant Coulomb repulsion
$U_{\alpha\alpha}$ between opposite-spin electrons on the
same level.

Ground state configurations and filling rules change when a magnetic field
is applied. It affects both single particle
energies and $U$- and $J$-integrals
through the induced changes in the wavefunction
localization.
For sufficiently large values, the magnetic field energetically favours
configurations with higher total angular quantum number $m$.
This is the physical origin of the wiggles in the
$A\!\left(N\right)$ vs.~$B$ plot shown in
Fig.~\ref{fig2} and observed in the experiments of Ref. 
\onlinecite{Tarucha}.

Figure \ref{fig3} shows the Coulomb and exchange integrals vs.~$B$ for the
first states, obtained for $\hbar\omega_0 =7.5$~meV.
For comparison, we also
show the corresponding quantities calculated within
the simplified 2D model.
We clearly see that $U$-integrals describing the interaction between
opposite-spin electrons are  smaller by a few meV
in the case of 3D confinement, while the
differences in the interaction between parallel-spin electrons
are much smaller.
This affects dramatically the energy balance which determines
ground state configurations, thus clearly showing the inadequacy
of a pure 2D description of state-of-the-art QD structures.

The proposed approach shares with the Hartree-Fock (HF) method the {\em
ansatz} for total energy, expressed as the average of the exact Hamiltonian
over a single Slater determinant; the variational prescription - allowing
the construction of optimal single particle orbitals through
self-consistency - is not contained in our approach. We notice however that
in the present case the results of perturbation theory are already much 
closer to the {\em exact} results than in previous 2D 
calculations:\cite{tobepublished} this is due to the fact that Coulomb
integrals are artificially enhanced in 2D with respect to the realistic 3D
description. Hence the role of self-consistency and the approximations of
HF theory are much less relevant in 3D. Finally, we point out that the
present formalism can be reduced to a Hubbard model description: expression
(\ref{eq:E}) is in fact formally equivalent to the exact solution of a
single-site Hubbard Hamiltonian. This is important in view of a direct 
extension of this approach to coupled dots. 

In summary, we have proposed an effective theoretical approach for the
analysis of addition spectra of quantum dots by combining 
a careful 3D description of electron-electron interaction with
the realistic treatment of single particle properties.
By construction, the proposed scheme is ideally suited for applications
to dots with confinement potentials of arbitrary shapes and symmetries.
A full 3D description of Coulomb interaction is found to be crucial
in determining single electron conductance properties of realistic 
nanostructures.

\bigskip
We are grateful to C.~Calandra for useful discussions.

%
%

%
%
%
\begin{figure}
\caption{Calculated addition energy increment
$\Delta A\!\left(N\right)$ vs.~electron
number $N$ for different QD
structures characterized by a parabolic potential 
in the $xy$ plane (confining energy $\hbar\omega_0$) and by a finite-barrier
quantum well potential along the $z$ direction.
}
\label{fig1}
\end{figure}
\begin{figure}
\caption{Addition energy $A\!\left(N\right)$
vs.~magnetic field calculated for a
realistic (3D) QD structure with confinement energy
$\hbar\omega_0 = 7.5$~meV and for different electron numbers.
Ground state configurations, varying with $B$, are also shown.
}
\label{fig2}
\end{figure}
\begin{figure}
\caption{Comparison between Coulomb and exchange integrals 
$U_{\alpha\beta}$
and $J_{\alpha\beta}$ calculated assuming a simple two dimensional
parabolic model (2D) and the corresponding three dimensional model (3D).
$\alpha$ and $\beta$ denote the sets of
radial and angular quantum numbers ($n,m$) for the two single particle states
involved in the interaction.
}
\label{fig3}
\end{figure}
\end{document}